\documentclass{article}
\usepackage{amssymb}
\usepackage{ismir,amsmath,cite,url}
\usepackage{graphicx}
\usepackage{color}
\usepackage{multirow}

\title{MidiNet: A Convolutional Generative Adversarial Network for Symbolic-domain Music Generation}

\oneauthor
 {Li-Chia Yang, Szu-Yu Chou, Yi-Hsuan Yang}
{Research Center for IT innovation, Academia Sinica, Taipei, Taiwan\\{\tt \{richard40148, fearofchou, yang\}@citi.sinica.edu.tw}}

\begin{document}

\maketitle
\begin{abstract}
Most existing neural network models for 
music generation use recurrent neural networks. 
However, the recent WaveNet model proposed by DeepMind 
shows that
convolutional neural networks (CNNs)
can also generate realistic musical waveforms in the audio domain. 
Following this light, we investigate using CNNs for generating melody (a series of MIDI notes) one bar after another in the symbolic domain. 
In addition to the generator, we use a discriminator to learn the distributions of melodies, making it a generative adversarial network (GAN). Moreover, we propose a novel conditional mechanism to exploit available prior knowledge, so that the model can generate melodies either from scratch, by following a chord sequence, or by conditioning on the melody of previous bars (e.g. a priming melody), among other possibilities. 
The resulting model, named MidiNet, can be expanded to generate music with multiple MIDI channels (i.e. tracks). 
We conduct a user study to compare the melody of eight-bar long generated by MidiNet and by Google's MelodyRNN models,  each time using the same priming melody. 
Result shows that MidiNet performs comparably with MelodyRNN models in being realistic and pleasant to listen to, yet MidiNet's melodies are reported to be much more interesting.
\end{abstract}

\section{Introduction}
\label{section:intro}


Algorithmic composition is not a new idea. 
The first computational model for algorithmic composition dates back to 1959 \cite{ExperimentalMusic}, according to the survey of Papadopoulos and Wiggins \cite{Papadopoulos99aimethods}.
People have also used (shallow) neural networks for music generation since 1989 \cite{todd89ann}.
It was, however, only until recent years when deep neural networks demonstrated their ability in learning from big data collections that generating music by neural networks became a trending topic.
Lots of deep neural network models 
for music generation
have been proposed just over the past two years \cite{crnn,DBach,Magenta,song_from_pi,wavenet,Mason,sampleRNN,fastWaveNet,AI_duet,waveNet2,sun16arxiv,jaquesGTE16}. 

%
%
The majority of existing neural network models for music generation use recurrent neural networks (RNNs) and their variants, presumably for music generation is inherently about generating sequences \cite{music_rnn, music_rnn2,todd89rnn,graves13}. 
These models differ in the model assumptions and the way musical events are represented and predicted, but they all use information from the previous events to condition the generation of the present one. Famous examples include the MelodyRNN models \cite{Magenta} for \emph{symbolic-domain} generation (i.e. generating MIDIs) and the SampleRNN model \cite{sampleRNN} for \emph{audio-domain} generation (i.e. generating WAVs).

%
%
Relatively fewer attempts have been made to use deep convolutional neural networks (CNNs) for music generation.
A notable exception is the WaveNet model \cite{wavenet} proposed recently for audio-domain generation. 
It generates one audio sample at a time, with the predictive distribution for each sample conditioned on  previous samples through dilated causal convolutions \cite{wavenet}. WaveNet shows it possible for CNNs to generate realistic music. 
This is encouraging, as CNNs are typically faster to train and more easily parallelizable than RNNs \cite{pixelCNN}. 

Following this light, we investigate in this paper a novel CNN-based model for symbolic-domain generation, focusing on melody generation.\footnote{In general, a melody may be defined as a succession of (monophonic) musical notes expressing a particular musical idea.} Instead of creating a melody sequence continuously, we propose to generate melodies \emph{one bar (measure) after another}, in a successive manner. This allows us to employ convolutions on a 2-D matrix representing the presence of notes over different time steps in a bar. 
We can have such a score-like representation for each bar for either a real or a generated MIDI.

%
%
\begin{table*}[t]
\centering
\caption{Comparison between recent neural network based music generation models}
\label{table: model}
\vspace{1mm}
\begin{tabular}{l|cccc|c|c}
     & MelodyRNN   & Song from PI & DeepBach & C-RNN-GAN & MidiNet  & WaveNet
\\ 
& \cite{Magenta} &   \cite{song_from_pi} & \cite{DBach} & \cite{crnn} & (this paper) & \cite{wavenet} 
\\
\hline
core model  & RNN    & RNN & RNN    & RNN  & CNN  & CNN \\
data type & symbolic  & symbolic  & symbolic  & symbolic & symbolic & audio       \\
genre specificity & --- & --- & Bach chorale  & ---    &  --- & --- \\
\hline
mandatory prior knowl- & priming     & music scale \&   & \multirow{2}{*}{---} & \multirow{2}{*}{---}    & \multirow{2}{*}{---}     & priming  \\
edge & melody & melody profile &   &  & & wave  \\
\hline
follow a priming melody    &  $\surd$   &  $\surd$  &      &     & $\surd$       &  $\surd$   \\
follow a chord sequence    &      &     &     &     & $\surd$          & \\
generate multi-track music &      & $\surd$   & $\surd$     &       & $\surd$             &  $\surd$ \\
use  GAN      &     &     &      & $\surd$   & $\surd$    &    \\
use versatile conditions    &    &    &     &    & $\surd$   & \\
open source code &   $\surd$   &     &    &   $\surd$   &   $\surd$    &  \\
\end{tabular}
\end{table*}

Moreover, to emulate \emph{creativity} \cite{Papadopoulos99aimethods} and encourage diverse generation result, we use random noises as input to our \emph{generator} CNN. The goal of the generator is to transform random noises into the aforementioned 2-D score-like representation, that ``appears'' to be from real MIDI. This transformation is achieved by a special convolution operator called transposed convolution \cite{cnn_arithmetic}. Meanwhile, we learn a \emph{discriminator} CNN that takes as input a 2-D score-like representation and predicts whether it is from a real or a generated MIDI, thereby informing the generator how to appear to be real. This amounts to a generative adversarial network (GAN) \cite{GAN,GANtutorial,DCGAN, cGAN2,ImprovedGAN}, which learns the generator and discriminator
iteratively under the concept of minimax two-player game theory. 

This GAN alone does not take into account the temporal dependencies across different bars. To address this issue, we propose a novel conditional mechanism to use music from the previous bars to condition the generation of the present bar. This is achieved by learning another CNN model, which we call the \emph{conditioner} CNN, to incorporate information from previous bars to intermediate layers of the generator CNN. This way, our model can ``look back'' without a recurrent unit as used in RNNs.
Like RNNs, our model can generate music of arbitrary number of bars.

Because we use random noises as inputs to our generator, our model can generate melodies \emph{from scratch}, i.e. without any other prior information. However, due to the conditioner CNN, our model has the capacity to exploit whatever prior knowledge that is available and can be represented as a matrix. For example, our model can generate music by following a chord progression, or by following a few starting notes (i.e. a priming melody). Given the same priming melody, our model can generate different results each time, again due to the random input.



The proposed model can be extended to generate different types of music, by using different conditions. 
Based on an idea called \emph{feature matching} \cite{ImprovedGAN}, we propose a way to control the influence of such conditions on the generation result.
We can then control, for example, how much the current bar should sound like the previous bars.
Moreover, our CNNs can be easily extended to deal with tensors instead of matrices,
to exploit multi-channel MIDIs and to generate music of multiple tracks or parts.
We believe such a highly adaptive and generic model structure can be a useful alternative to RNN-based designs. We refer to this new model as the MidiNet.

In our experiment, we conduct a
user study to compare the melodies generated by MidiNet and MelodyRNN models \cite{Magenta}. For fair comparison, 
we use the same priming melodies for them to generate melodies of eight-bar long (including the primers), without any other prior information. 
To demonstrate the flexibility of MidiNet, we provide the result of two additional settings: one uses additionally chord progressions of eight-bar long to condition the generation, and the other uses a slightly different network architecture to generate more creative music. For reproducibility, the source code and pre-trained models of MidiNet are released online\footnote{\url{https://github.com/RichardYang40148/MidiNet}}.

\begin{figure*}
\centering
\includegraphics[width=\linewidth]{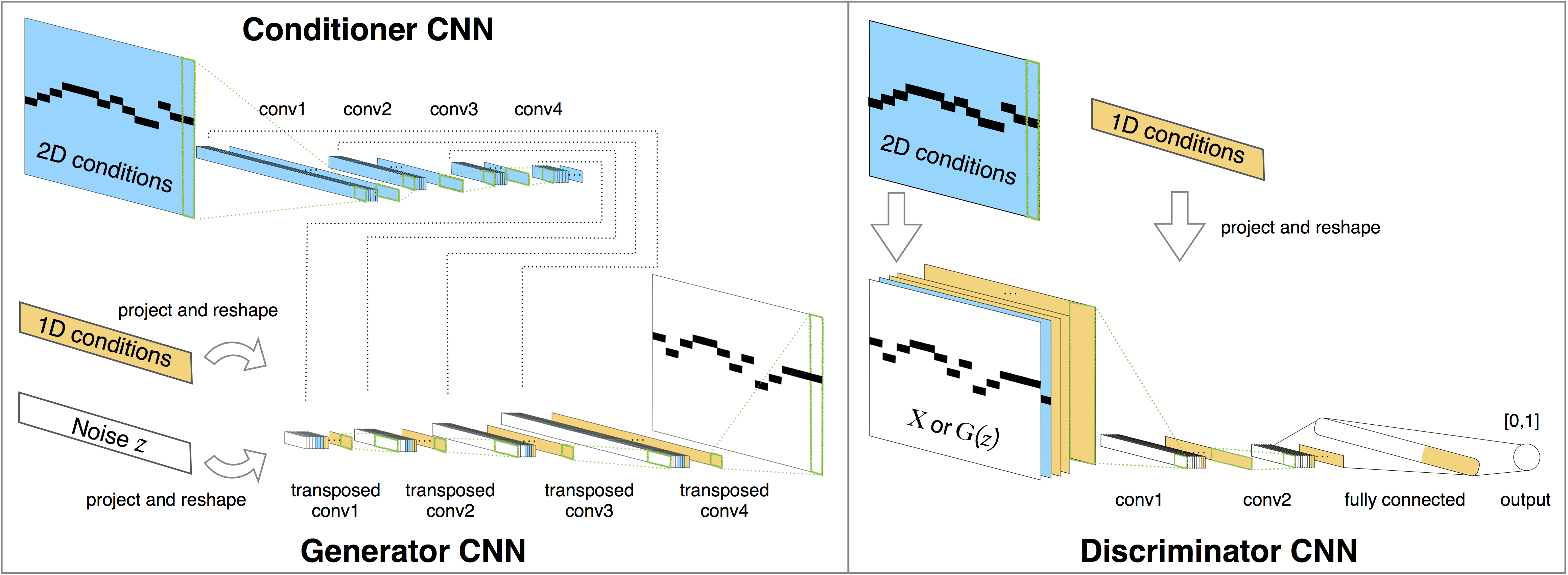}
\caption{System diagram of the proposed MidiNet model for symbolic-domain music generation.}
\label{fig: Architecture}
\end{figure*}

\section{Related Work}
\label{sec:related}


A large number of deep neural network models have been proposed lately for music generation.
This includes models for generating a melody sequence or audio waveforms by following a few priming notes \cite{Magenta, jaquesGTE16, wavenet,waveNet2,sampleRNN,fastWaveNet}, accompanying a melody sequence with music of other parts \cite{DBach}, or playing a duet with human \cite{Mason,AI_duet}.

Table \ref{table: model} compares MidiNet  with a number of major related models. We briefly describe each of them below.

The MelodyRNN models \cite{Magenta} proposed by the Magenta Project from the Google Brain team are possibly among the most famous examples of symbolic-domain music generation by neural networks. In total three RNN-based models were proposed, including two variants that aim to learn longer-term structures, the lookback RNN and the attention RNN  \cite{Magenta}. Source code and pre-trained models for the three models are all publicly available.\footnote{\url{https://github.com/tensorflow/magenta/tree/master/magenta/models/melody_rnn} (accessed 2017-4-26)} As the main function of MelodyRNN is to generate a melody sequence from a priming melody, we use the MelodyRNN models as the baseline in our evaluation.

Song from PI \cite{song_from_pi} is a hierarchical RNN model that uses a hierarchy of recurrent layers to generate not only the melody but also the drums and chords, leading to a multi-track pop song. This model nicely demonstrates the ability of RNNs in generating multiple sequences simultaneously. However, it requires prior knowledge of the musical scale and some profiles of the melody to be generated \cite{song_from_pi}, which is not needed in many other models, including 
MidiNet.

DeepBach \cite{DBach}, proposed by Sony CSL, is specifically designed for composing polyphonic four-part chorale music in the style of J.~S. Bach. It is an RNN-based model that allows enforcing user-defined constraints such as rhythm, notes, parts, chords and cadences.

C-RNN-GAN \cite{crnn} is to date the only existing model that uses GAN for music generation, to our best knowledge. It also takes random noises as input as MidiNet does, to generate diverse melodies. However, it lacks a conditional mechanism \cite{reed2016generative, isola2016image,cGAN } to generate music by following either a priming melody or a chord sequence.

WaveNet \cite{wavenet,waveNet2} is a CNN-based model proposed by DeepMind for creating raw waveforms of speech and music. 
The advantage of audio-domain generation is the possibility of creating new sounds, but we choose to focus on symbolic-domain generation in this paper.

\section{Methods}
\label{sec:method}

A system diagram of MidiNet is shown in Figure \ref{fig: Architecture}.
Below, we present the technical details of each major component. 


\subsection{Symbolic Representation for Convolution}

Our model uses a symbolic representation of music in fixed time length, by dividing a MIDI file into bars. The note events of a MIDI channel can be represented by an $h$-by-$w$ real-valued matrix $\mathbf{X}$,
where $h$ denotes the number of MIDI notes we consider, possibly including one more dimension for representing silence, 
and $w$ represents the number of time steps we use in a bar.
For melody generation, there is at most one active note per time step. We use a binary matrix $\mathbf{X} \in \{0,1\}^{h \times w}$ if we omit the velocity (volume) of the note events. We use multiple matrices per bar if we want to generate multi-track music.

In this representation, we may not be able to easily distinguish between a long note and two short repeating notes (i.e. consecutive notes with the same pitch).
Future extensions can be done to emphasize the note onsets. 




\subsection{Generator CNN and Discriminator CNN}

The core of MidiNet is a modified deep convolutional generative adversarial network (DCGAN) \cite{DCGAN}, which aims at learning a discriminator $D$ to distinguish between real (authentic) and generated (artificial) data, and a generator $G$ that ``fools'' the discriminator. As typical in GANs, the input of $G$ is a vector of random noises $\mathbf{z}\in\mathbb{R}^l$, whereas the output of $G$ is an $h$-by-$w$ matrix $\widehat{\mathbf{X}}=G(\mathbf{z})$ that ``appears'' to be real to $D$.
GANs learn $G$ and $D$ by solving:
\begin{equation} 
\begin{split}
\underset{G}{\min}\underset{D}{\max}~V(D,G)=\mathbb{E}_{\mathbf{X}\sim p_\text{data}(\mathbf{X})}[\log(D(\mathbf{X}))]+\\
\mathbb{E}_{\mathbf{z}\sim p_z(\mathbf{z})}[\log(1-D(G(\mathbf{z})))]\,,
\end{split}
\label{eq:gan}
\end{equation} 
where $\mathbf{X}\sim p_\text{data}(\mathbf{X})$ denotes the operation of sampling from real data, and $\mathbf{z}\sim p_z(\mathbf{z})$ the sampling from a random distribution.
As typical in GANs, we need to train $G$ and $D$ iteratively multiple times, to gradually make a better $G$.

Our discriminator  
is a typical CNN with a few convolution layers, 
followed by fully-connected layers. 
These layers are optimized with a cross-entropy loss function, such that the output of $D$ is close to 1 for real data (i.e. $\mathbf{X}$) and 0 for those generated (i.e. $G(\mathbf{z})$). We use a sigmoid neuron at the output layer of $D$ so its output is in [0,1].

The goal of the generator CNN, on the other hand, is to make the output of $D$ close to 1 for the generated data. For generation, it has to transform a vector $\mathbf{z}$ into a matrix $\widehat{\mathbf{X}}$. This is achieved by using a few fully connected layers first, and then a few transposed convolution layers \cite{cnn_arithmetic} that ``upsamples'' smaller vectors/matrices into larger ones.

Owing to the nature of minimax games, the training of GANs is subject to issues of instability and mode collapsing \cite{GANtutorial}. 
Among the various possible techniques to improve the training of GANs \cite{ImprovedGAN, infoGAN, WGAN}, 
we employ the so-called feature matching and one-sided label smoothing \cite{ImprovedGAN} in our model.
The idea of feature matching is to add additional L2 regularizers to Eq. \ref{eq:gan}, such that the distributions of real and generated data are enforced to be close. Specifically, we add the following two terms when we learn $G$:
\begin{equation} 
\begin{split}
\lambda_{1} \| \mathbb{E}~\mathbf{X}-\mathbb{E}~G(\mathbf{z}) \|^2_2 +
\lambda_{2} \| \mathbb{E}~f(\mathbf{X})-\mathbb{E}~f(G(\mathbf{z})) \|^2_2\,,
\end{split}
\label{eq:loss}
\end{equation} 
where $f$ denotes the first convolution layer of the discriminator, and
$\lambda_1$, $\lambda_2$ are parameters to be set empirically.


\subsection{Conditioner CNN}
\label{sec:method:conditioner}

In GAN-based image generation, people often use a vector to encode available prior knowledge that can be used to condition the generation. This is achieved by reshaping the vector and then adding it to different layers of $G$ and $D$, to provide additional input \cite{cGAN}. Assuming that the conditioning vector has length $n$, to add it to an intermediate layer of shape $a$-by-$b$ we can duplicate the values $ab$ times to get a tensor of shape $a$-by-$b$-by-$n$, and then concatenate it with the intermediate layer in the feature map axis. This is illustrated by the light orange blocks in Figure \ref{fig: Architecture}. We call such a conditional vector \emph{1-D conditions}.

As the generation result of our GAN is an $h$-by-$w$ matrix of notes and time steps, it is convenient if we can perform conditioning directly on each entry of the matrix. For example, the melody of a previous bar can be represented as another $h$-by-$w$ matrix and used to condition the generation of the present bar. We can have multiple such matrices to learn from multiple previous bars. We can directly add such a conditional matrix to the input layer of $D$ to influence all the subsequent layers. However, to exploit such \emph{2-D conditions} in $G$, we need a mechanism to reshape the conditional matrix to smaller vectors of different shapes, to include them to different intermediate layers of $G$.

We propose to achieve this by using a conditioner CNN
that can be viewed as a \emph{reverse} of the generator CNN. 
As the blue blocks in Figure \ref{fig: Architecture} illustrates, the conditioner CNN uses a few convolution layers to process the input $h$-by-$w$ conditional matrix. 
The conditioner and generator CNNs use exactly the same filter shapes in their convolution layers, so that the outputs of their convolution layers have ``compatible'' shapes.
In this way, we can concatenate the output of a convolution layer of the conditioner CNN to the input of a corresponding transposed convolution layer of the generator CNN, to influence the generation process.
In the training stage, the conditioner and generator CNNs are trained simultaneously, by sharing the same gradients.

\subsection{Tunning for Creativity}
\label{sec:method:opt2}

We propose two methods to control the trade-off between creativity and discipline of MidiNet. The first method is to manipulate the effect of the conditions by using them only in part of the intermediate transposed convolution layers of $G$, to give $G$ more freedom from the imposed conditions.
The second method capitalizes the effect of the feature matching technique \cite{ImprovedGAN}: we can increase the values of $\lambda_1$ and $\lambda_2$ to make the generated music sounds closer to existing music (i.e. those observed in the training set).

\section{Implementation}
\label{sec:data}
\subsection{Dataset} 

As the major task considered in this paper is melody generation, for training MidiNet we need a MIDI dataset that clearly specifies per file which channel corresponds to the melody. To this end, we crawled a collection of 1,022 MIDI tabs of pop music from TheoryTab,\footnote{\url{https://www.hooktheory.com/theorytab}}
which provides exactly two channels per tab, one for melody and the other for the underlying chord progression. With this dataset, we can implement at least two versions of MidiNets: one that learns from only the melody channel for fair comparison with MelodyRNN \cite{Magenta}, which does not use chords, and the other that additionally uses chords to condition melody generation, to test the capacity of MidiNet. 

For simplicity, we filtered out MIDI tabs that contain chords other than the 24 basic chord triads (12 major and 12 minor chords). Next, we segmented the remaining tabs every 8 bars, and then pre-processed the melody channel and the chord channel separately, as described below.

For melodies, we fixed the smallest note unit to be the sixteenth note, making $w=16$. Specifically, we prolonged notes which have a pause note after them. If the first note of a bar is a pause, we extended the second note to have it played while the bar begins. There are other exceptions such as triplets and shorter notes (e.g. 32nd notes), but we chose to exclude them in this implementation. Moreover, for simplicity, we shifted all the melodies into two octaves, from \texttt{C4} to \texttt{B5}, and neglected the velocity of the note events. Although our melodies would use only 24 possible notes after these preprocessing steps, we considered all the 128 MIDI notes (i.e. from \texttt{C0} to \texttt{G10}) in our symbolic representation. In doing so, we can detect model collapsing \cite{GANtutorial} more easily, by checking whether the model generates notes outside these octaves. As there are no pauses in our data after preprocessing, we do not need a dimension for silence. Therefore, $h=128$.

For chords, instead of using a 24-dimensional one-hot vector, we found it more efficient to use a chord representation that has only 13 dimensions--- the first 12 dimensions for marking the key, 
and the last for the chord type (i.e. major or minor), 
as illustrated in Table \ref{tab:chord}.
We pruned the chords such that there is only one chord per bar.

\begin{table}
\centering
\caption{13-dimensional chord representation}
\begin{tabular}{l|c|l}
 & dimensions 1--12 & 13 \\ \hline
major & C, C\#, D, D\#, E, F, F\#, G, G\#, A, A\#, B  & 0  \\
minor & A, A\#, B, C, C\#, D, D\#, E, F, F\#, G, G\# & 1 
\label{tab:chord}
\end{tabular}
\end{table}

After these preprocessing steps, we were left with 526 MIDI tabs (i.e. 4,208 bars).\footnote{In contrast, MelodyRNN models \cite{Magenta} were trained on \emph{thousands} of MIDI files, though the exact number is not yet disclosed.}
For data augmentation, we circularly shifted 
the melodies and chords to any of the 12 keys in equal temperament, leading to a final dataset of 50,496 bars of melody and chord pairs for  training.

\subsection{Network Specification}
\label{subsection:training}

Our model was implemented in TensorFlow.
For the generator, we used as input random vectors of white Gaussian noise of length $l=100$. Each random vector go through two fully-connected layers, with 1024 and 512 neurons respectively, before being reshaped into a 1-by-2 matrix. We then used four transposed convolution layers: the first three use filters of shape 1-by-2 and two strides \cite{cnn_arithmetic}, and the last layer uses filters of shape 128-by-1 and one stride. 
Accordingly, our conditioner has four convolution layers, which use 128-by-1 filters for the first layer, and 1-by-2 filters for the other three.
For creating a monophonic note sequence, we added a layer to the end of $G$ to turn off per time step all but the note with the highest activation.



As typical in GANs, the discriminator is likely to overpower the generator, leading to the so-called vanishing gradient problem \cite{GANtutorial,WGAN}.
We adopted two strategies to weaken the discriminator.
First, in each iteration, we updated the generator and conditioner twice, but the discriminator only once. 
Second, we used only two convolution layers (14 filters of shape 128-by-2, two strides, and 77 filters of shape 1-by-4, two strides) and one fully-connected layer (1,024 neurons) for the discriminator. 

We fine-tuned the other parameters of MidiNet and considered the following three variants in our experiment.
 
\subsubsection{Model 1: Melody generator, no chord condition}
\label{sec:impl:model1}

This variant uses the melody of the previous bar to condition the generation of the present bar. We used this 2-D condition in all the four transposed convolution layers of $G$. We set the number of filters in all the four transposed convolution layers of $G$ and the four convolution layers of the conditioner CNN to 256. The feature matching parameters $\lambda_{1}$ and $\lambda_{2}$ are set to 0.1 and 1, respectively. We did not use the 2-D condition for $D$, requiring it to distinguish between real and generated melodies from the present bar.

In the training stage, we firstly added one empty bar before all the MIDI tabs, and then randomly sampled two consecutive bars from any tab. We used the former bar as an instance of real data (i.e. X) and the input to $D$, and the former bar (which is a real melody or all zeros) as a 2-D condition and the input to the conditioner CNN
Once the model was trained, we used $G$ to generate melodies of 8-bar long in the following way: the first bar was composed of a real, priming melody sampled from our dataset; the generation of the second bar was made by $G$, conditioned by this real melody; starting from the third bar, $G$ had to use the (artificial) melody it generated previously for the last bar as the 2-D condition. This process repeated until we had all the eight bars.\footnote{It is also possible to use multiple previous bars to condition our generation, but we leave this as a future extension.}



\subsubsection{Model 2: Melody generator with chord condition, stable mode}
\label{sec:impl:model2}

This variant additionally uses the chord channel. Because our MIDI tabs use one chord per bar, we used the chord (a 13-dimensional vector; see Table \ref{tab:chord}) of the present bar as a 1-D condition for generating the melody for the same bar. We can say that our model is generating a melody sequence that fits the given chord progression.

To highlight the chord condition, we used the 2-D previous-bar condition only in the last transposed convolution layer of $G$. In contrast, we used the 1-D chord condition in all the four transposed convolution layer of $G$, as well as the input layer for $D$. Moreover, we set $\lambda_1=0.01$, $\lambda_2=0.1$, and used 128 filters in the transposed convolution layers of $G$ and only 16 filters in the convolution layers of the conditioner CNN.
As a result, 
the melody generator is more \emph{chord-dominant} and \emph{stable}, for it would mostly follow the chord progression and seldom generate notes that  violate the constraint imposed by the chords. 

\subsubsection{Model 3: Melody generator with chord condition, creative mode}
\label{sec:impl:model3}

This variant realizes a slightly more creative melody generator by placing the 2-D condition in every transposed convolution layer of $G$. In this way, $G$ would sometimes violate the constraint imposed by the chords, to somehow adhere to the melody of the previous bar. Such violations sometimes sound unpleasant, but can be sometimes creative. Unlike the previous two variants, 
we need to listen to several melodies generated by this model to handpick good ones. However, we believe such a model can still be useful for assisting and inspiring human composers. 

\section{Experimental Result}
\label{sec:exp}

\begin{figure}
\centering
\includegraphics[width=\linewidth]{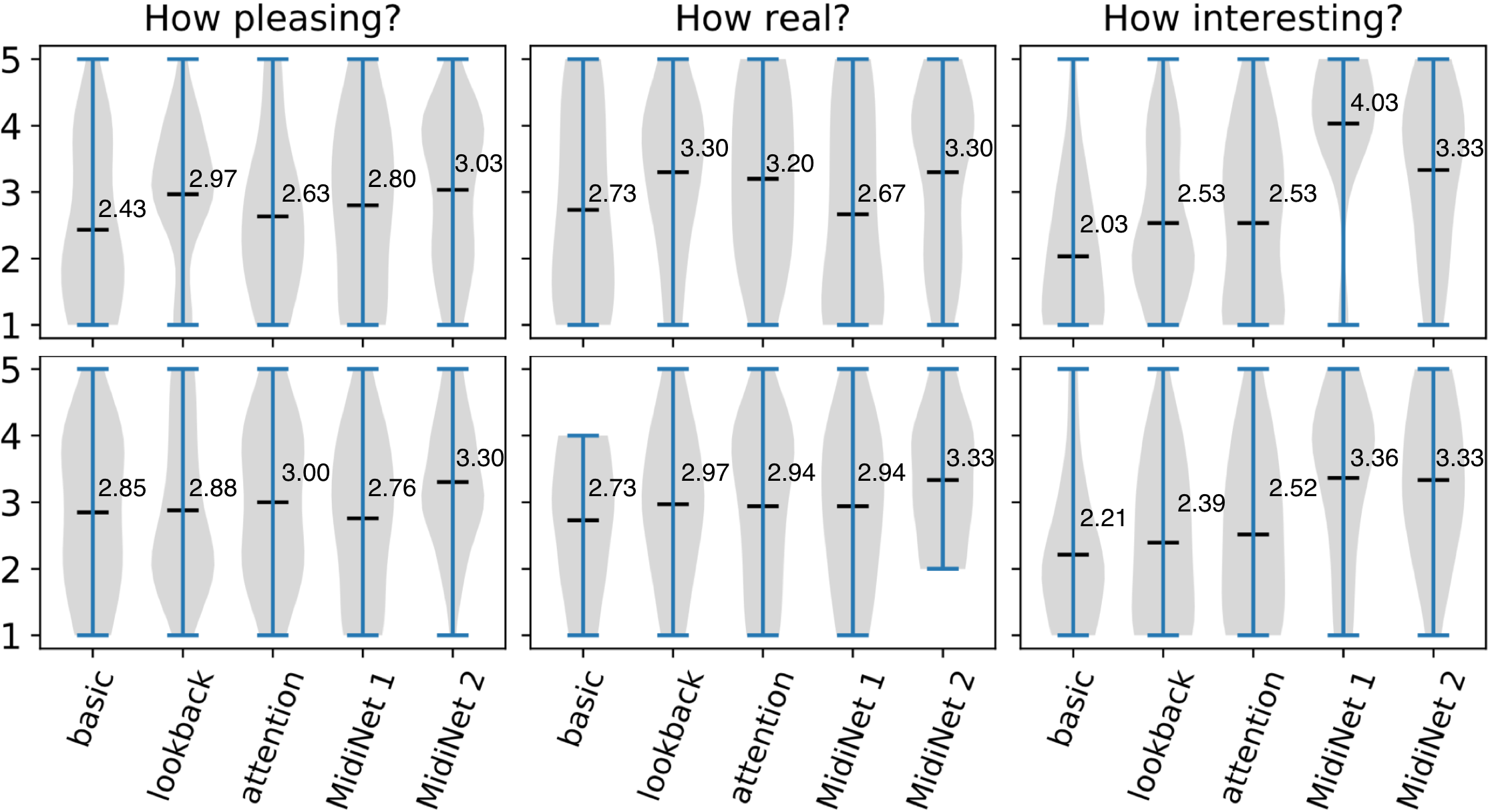}
\caption{Result of a user study comparing MelodyRNN and MidiNet models, for people (top row) with musical backgrounds and (bottom) without musical backgrounds. The middle bars indicate the mean values. Please note that MidiNet Model 2 takes the chord condition as additional information.}
\label{fig: user}
\end{figure}

\begin{figure*}
\centering
\includegraphics[width=\linewidth]{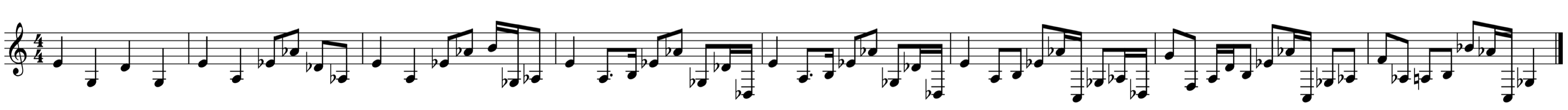}
\\(a) MidiNet model 1\\
\includegraphics[width=\linewidth]{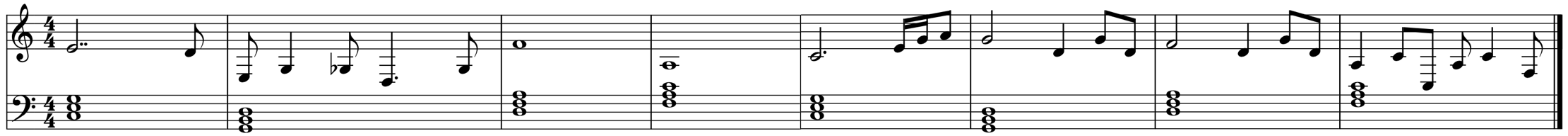}
\\(b) MidiNet model 2\\
\includegraphics[width=\linewidth]{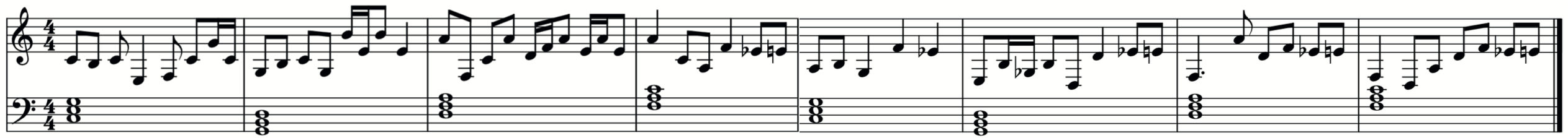}
\\(c) MidiNet model 3
\caption{Example result of the melodies (of 8 bars) generated by different implementations of MidiNet.}
\label{fig: Sample}
\end{figure*}

To evaluate the aesthetic quality of the generation result, a user study that involves human listeners is needed. We conducted a study with 21 participants. Ten of them understand basic music theory and have the experience of being an amateur musician, so we considered them as people with musical backgrounds, or \emph{professionals} for short.

We compared MidiNet with three MelodyRNN models pre-trained and released by Google Magenta: the basic RNN, the lookback RNN, and the attention RNN \cite{Magenta}. We randomly picked 100 priming melodies from the training data \footnote{Even though these priming melodies are in the training data, MidiNet generates melodies that are quite different from the existing ones.} 
and asked the models create melodies of eight bars by following these primers. We considered two variants of MidiNet in the user study: model 1 (Section \ref{sec:impl:model1}) for fair comparison with MelodyRNN, and model 2 (Section \ref{sec:impl:model2}) for probing the effects of using chords. Although the result of model 2 was generated by additionally following the chords, we did not playback the chord channel in the user study.

We randomly selected the generation result of three out of the 100 priming melodies for each participant to listen to, leading to three sets of music.
To avoid bias, we randomly shuffled the generation result by the five considered models, such that in each set the ordering of the five models is different. 
The participants were asked to stay in a quiet room separately and used a headphone for music listening through the Internet, one set at a time.
We told them that some of the music ``might be'' real, and some might be generated by machine, although all of them were actually automatically generated.
They were asked to rate the generated melodies in terms of the following three metrics: \emph{how pleasing}, \emph{how real}, and \emph{how interesting}, from 1 (low) to 5 (high) in a five-point Likert scale. 

The result of the user study is shown in Figure \ref{fig: user} as violin plots, where the following observations can be made.
First, among the MelodyRNN models, lookback RNN and attention RNN consistently outperform basic RNN across the three metrics and two user groups (i.e. people with and without musical backgrounds), which is expected according to the report of Magenta \cite{Magenta}. The mean values for lookback RNN are around 3 (medium) for being pleasant and realistic, and around 2.5 for being interesting.

Second, MidiNet model 1, which uses only the previous bar condition, obtains similar ratings as the MelodyRNN models in being pleasant and realistic. This is encouraging, as MelodyRNN models can virtually exploit all the previous bars in generation.
This result demonstrates the effectiveness of the proposed conditioner CNN in learning temporal information.
Furthermore, we note that the melodies generated by MidiNet model 1 were found much more interesting than those generated by MelodyRNN. The mean value in being interesting is around 4 for people with musical backgrounds, and 3.4 for people without musical backgrounds. The violin plot indicates that the ratings of the professionals are mostly larger than 3.

Third, MidiNet model 2, which further uses chords, obtains the highest mean ratings in being pleasant and realistic for both user groups. 
In terms of interestingness, it also outperforms the three MelodyRNN models, but is inferior to MidiNet model 1, especially for professionals.


According to the feedback from the professionals, a melody sounds artificial if it lacks variety or violates principals in (Western) music theory. 
The result of MidiNet model 1 can  sound artificial, for it relies on only the previous bar and hence occasionally generates ``unexpected'' notes. In contrast, the chord channel provides a musical context that can be effectively used by MidiNet model 2 through the conditional mechanism. However, occasional violation of music theory might be a source of interestingness and thereby creativity. 
For example, the professionals reported that the melodies generated by MelodyRNN models are sometimes too repetitive, or ``safe,'' making them artificial and less interesting.
It might be possible to further fine tune our model to reach a better balance between being real and being interesting, but we believe our user study has shown the promise of MidiNet.

Figure \ref{fig: Sample} shows some melodies generated by different implementations of MidiNet, which may provide insights into MidiNet's performance.
Figure \ref{fig: Sample}(a) shows that MidiNet model 1 can effectively exploit the previous bar condition---most bars start with exactly the same first two notes (as the priming bar) and they use similar notes in between.
Figure \ref{fig: Sample}(b) shows the result of MidiNet model 2, which highlights the chord condition.
Figure \ref{fig: Sample}(c) shows that MidiNet can generate more creative result by making the chord condition and previous bar condition equally strong. We can see stronger connections between adjacent bars from the result of this MidiNet model 3.
For more audio examples, please go to  \url{https://soundcloud.com/vgtsv6jf5fwq/sets
}.



\section{Conclusion}
\label{sec:concl}


We have presented MidiNet, 
a novel CNN-GAN based model for MIDI generation. It has a conditional mechanism to exploit versatile prior knowledge of music. It also has a flexible architecture and can generate different types of music depending on input and specifications. Our evaluation shows that it can be a powerful alternative to RNNs.




For future work, we would extend MidiNet to generate multi-track music, to include velocity and pauses by training the model by using richer and larger MIDI data. We are also interested in using ideas of reinforcement learning \cite{RL} to incorporate principles of music theory \cite{jaquesGTE16}, and to take input from music information retrieval models such as genre recognition \cite{choiFSC16} and emotion recognition \cite{yang11book}. 



\bibliography{ref}
\end{document}